# Information Seeking and Information Processing Behaviors Among Type 2 Diabetics

Sarah Masud Preum, MCS, Kate Clark, MSN RN APHN, Ashley Davis, MPH
Konstantine Khutsishvili, MD, Rupa S Valdez, PhD
University of Virginia, Charlottesville, VA

**Abstract**

*This study sought to assess self-education behaviors of type 2 diabetics to understand more about how these behaviors affect the self-management of this common chronic disease. We focused on two factors of self-education behavior; information seeking and information processing. Using interviews and qualitative analysis techniques, several themes emerged: Patients are often dissatisfied with information from "official" sources, have difficulty evaluating the trustworthiness of information sources, and desire information that is more personally relevant to them. To address these concerns we propose a web-based platform that can assist type 2 diabetics in seeking personalized trustworthy information and thus make information processing and utilization easier for them.*

**Introduction**

Type 2 Diabetes Mellitus (T2DM) is a major public health concern in the United States. Over 29 million people were living with T2DM in 2012,[1] and T2DM accounts for nearly 14% of U.S. health care expenditures, with half of those costs being related to long term complications such as renal disease, retinopathy, foot ulcers and stroke.[2] Diabetes patients spend over 8,000 hours per year self-managing their condition outside the medical setting.[3] Patients' knowledge of their disease, ability to understand and comply with recommended lifestyle modifications and interpretation of medication regimes are vital components of self-management.[4,5] Proper education with respect to DM management has been shown to be cost effective, as appropriate preventive management can lead to less spending on drastic consequences in the future.[6] The American Diabetes Association's 2015 standards for care also underscore the importance of self-management education, as effective education leads to clinical improvement.[5] This study focuses on two important aspects of self-education in T2DM patients, namely, information seeking and information processing.

The health information seeking behavior of diabetics has been studied in variety of countries in various contexts. More than one study corroborated that more active information seeking was associated with better adherence to treatment regimens and other self-management activities.[7] While older populations tend to rely more on doctors and interpersonal relationships, younger patients rely more on online resources.[7,8] Older patients prefer to receive information passively from healthcare providers, believing that providers will tell them everything they need to know and direct them to appropriate resources.[9] Many of those same patients do not feel confident that they are able to personally locate adequate information on their own,[10] and have high levels of trust in their healthcare providers.[10,11]

In spite of these barriers, many patients do use the Internet for health information. The diabetes online community, which includes anyone engaged in online activity related to DM (blogging, searching for information, etc.), is growing. Fifty-one percent of patients living with one or more chronic conditions use the Internet to obtain health information,[12] and persons with chronic disease are more likely to discuss health information found online with a provider.[3] The growth of the online community and the rise of tools such as tailored social media websites and smartphone health apps are making it easier and easier to access information. T2DM patients who use the Internet for health information tend to be younger in age, have higher educational attainment, higher incomes, newer diagnoses, higher BMI, poor glycemic control and higher wait times to see a doctor.[13] The most common subjects researched on the internet by DM patients is weight loss and weight control.[12] Although the Internet provides a wealth of knowledge there are some caveats. One study found that accurate information was only provided 40% of time, because health information websites are not always moderated by professionals.[3] Therefore, processing this information often requires assistance from nurse practitioners and doctors.[9]

Information processing behaviors of T2DM patients are rarely directly explored in existing research. Some studies focusing on information seeking behavior of T2DM patients address the work of information processing and report that people sometimes find information too hard to understand.[14] Some barriers to information processing have been identified, such as an overwhelming volume of available information, conflicting information from different sources, and complexity of information. These barriers cause some patients to get information from their professional healthcare providers instead of the Internet.[9] In such cases, health care professionals can help to clarify confusion and guide patients as they identify credible sources of information.[9] To our knowledge, there are no other overt references to information processing in the T2DM population. The limited information about this important component of self-education reflects a gap in the current literature and deserves further examination.

Despite evidence demonstrating the significant role health education plays in T2DM patients' self-management, there is a need for more clarity in understanding specific self-education behaviors like information seeking and processing behaviors of T2DM patients and their impacts on T2DM management. Specifically, little is known about how and why patients prioritize and trust one source of health information over another. Learning more about T2DM patients' experience and preferences in disease related information seeking and processing is important in designing comprehensible, accessible, and accurate information sources to help patients improve their disease management.

**Methods**

The purpose of this study was to understand how individuals with T2DM search for, understand, and process information related to their disease. This qualitative study was part of a human factors design course project at the University of Virginia. The research team was comprised of an interdisciplinary team of four graduate students. This study was approved by the University of Virginia Institutional Review Board for Social and Behavioral Sciences.

**Recruitment.** Recruitment occurred through Facebook groups and pages created for individuals with T2DM. This strategy for participant recruitment has been documented as well suited for small, qualitative studies.[15] Data collection occurred through telephone or Skype interviews, depending on participant preference. Recruitment initially focused on English-speakers diagnosed with T2DM within the past two years. The language requirement was due to the fact that the study members speak English. Initially the study aimed at capturing the information seeking and processing experiences of recently diagnosed diabetics due to an assumption that this subgroup might have unique information needs and be more motivated to learn about their new diagnosis. Due to difficulties recruiting enough participants, the eligibility criteria was expanded to any interested English speaking person with a diagnosis of T2DM.

Recruitment took place from the end of September through the end of October 2016. Initially, the study team identified and contacted twenty-four Facebook groups related to T2DM. When contacting groups, members of the research team provided a brief explanation of the study, a link to the eligibility survey, and attached a flyer with further details.

Facebook groups can be labeled as either closed groups, requiring administrator approval to join, or open groups, where anyone with a Facebook account can join. Of the twenty-four groups contacted, six were open groups. For open groups, one study team member joined the group and then posted information about the study to the group's page. A private message was also sent to the administrator, asking them to share the study information with their members. For closed Facebook groups, a study team member contacted all group administrators initially by private message, shared the study information and link to the eligibility survey, and asked if they would post the information to their page for their members. Initially, two group administrators agreed to post the study information and flyer.

Due to a low response from administrators, study group members requested permission to join the closed groups, then contacted group administrators as a group member with the same request to post information for the study. This strategy was a response to Facebook's policy of not displaying any message notification if the message comes from an unknown user and sending the message to a separate message folder instead of the regular message folder. As a result, administrators may have overlooked the private messages being sent before the study group member joined their Facebook group. A total of six groups accepted our request to join and five of them allowed the study information and flyer to be posted in the group. In addition to groups, sixteen Facebook pages related to T2DM were contacted with the study information and flyer. But the study team received very low response rate. A potential reason of that is, as pointed out by a group administrator, it can take some time for a newly diagnosed diabetic to move past the shock of being diagnosed begin to seek out information online. After continued poor response rates for the eligibility survey, the study team obtained IRB approval to post the study information and flyer to their personal Facebook pages as well as to directly email personal acquaintances with ties to T2DM communities who could share the study information on behalf of the study team.

The sample was obtained through a convenience sampling method. Individuals interested in participating completed an eligibility survey consisting of eight questions. The survey, administered through surveymonkey.com, asked for contact information, demographic information, length of diagnosis and communication preferences. Individuals were contacted for interviews in the order in which they completed the survey. All individuals who completed the survey and responded to a follow-up email to arrange an interview were interviewed and included in the study results.

A total of sixteen survey responses were obtained. Only fifteen survey respondents were contacted to schedule interviews as one survey respondent did not provide any contact information. Of the fifteen respondents, seven did not respond to multiple attempts at contact via email or phone. The remaining eight respondents were recruited for the study. Study participants were compensated with a twenty-dollar gift card following completion of the interview. All participants provided verbal consent to participate in the study after being read the IRB-approved consent form.

**Data Collection and Analysis.** [15]Data collection occurred through telephone or Skype interviews, depending on participant preference. Interviews were conducted using an interview guide created by the study team. The guide was comprised of fifteen primary questions with probes as needed. The guide was divided into two parts. The first part of the guide included questions related to participants' personal experience with managing their diabetes. The second part of the guide focused on the participants' experiences with information seeking, processing and their impacts on T2DM management. It should be noted that this study was not designed with a particular theoretical framework in mind as to allow the study to progress without the constraints of a prior perspective. The data analysis process used an inductive approach of openly allowing themes to emerge from the data through review by the study team.

Seven interviews were conducted over the phone and one interview was conducted over Skype using the audio option. Interviews lasted between twenty minutes and one hour (average length was forty-five minutes) depending on how much the participants elaborated on their answers. Every interview was attended by two study team members, with one conducting the interview and the other listening as an observer. All interviews were recorded, the audio recording was immediately uploaded to a secure server and then deleted off of the recording device. Following each interview, the study group member who conducted the interview transcribed the audio. The study group member who listened in during the interview reviewed the transcript for accuracy. Transcripts were then uploaded into the coding software program, QSR NVivo 11. A qualitative content analysis of the data was performed to identify themes and categories to describe the data.

An iterative coding process was used to identify themes that emerged from the data. Two transcripts were reviewed and coded by each team member. Each member noted general themes or trends. Then, the study team came together to share and discuss what themes they had identified from the two transcripts and overall impressions. An initial coding structure was agreed upon. The remaining transcripts were each reviewed and coded by two team members, one who had participated in the interview and one who had not for added rigor. The study team remained in communication throughout the data analysis portion of the study in order to share ideas, concerns and strategies. The final coding structure included five primary themes and nine categories.

**Results**

The final study sample included eight participants. Seven identified themselves as female and one as male. Study participants were mostly older adults with three individuals who were fifty-one to sixty and two who were over sixty years old. Two participants were between the ages of forty-one and fifty and one participant was between eighteen and thirty years old. The length of time that the participants had been diagnosed with T2DM varied from three months to twelve years with the average length of diagnosis for the sample being three years ten months.

**Themes.** Overall, study participants described their experiences with obtaining and managing information related to their diagnosis of T2DM. Through data analysis, five key themes were identified: information seeking, information processing, information utilization, trust, and personalization.

*Information Seeking.* This theme captured different aspects of information seeking behavior, such as sources of information and their perceived value, motivation for information seeking, frequency and topics researched. Participants reported a wide range of information sources, including support groups, websites, social media, magazines, books, newsletters, social contacts, diabetes education classes, and health care providers. With respect to health care providers, some participants appreciated that their doctors took the time to make sure their questions were answered, whereas other participants were not satisfied with the amount of information given to them by healthcare providers after their diagnosis. One participant said, "I was given very little information at diagnosis. I was handed a prescription for insulin and metformin and a meter. Nobody told me how to use it, [or] what to do."

Six out of eight participants participated in diabetes education classes after their diagnosis. Diabetes education programs were reported to be generally ineffective, due to a lack of personalization, brevity of presentation, and inconsistent or conflicting information. In the words of one participant, "Most diabetic educators, because of guidelines that they have to follow, don't recommend [my current low carb, high fat diet], and…with the information I would have gotten at a diabetic educator class, I would not have had the successful results that I've had so far."

Motivation to seek out information is another important component of information seeking behavior, as it influenced how often participants looked for information and what topics they looked up. Participants mentioned different factors that motivated them to seek information, such as the desire to control blood sugar, getting inadequate information from their healthcare providers, and physiological events (e.g., feeling discomfort, having a spike in blood sugar). "I wake up and my blood sugar, my fasting number is higher than it normally is. And I'm thinking, 'Why on earth does this happen?' Other people may have ideas about why that would happen." Although some participants reported actively looking for information on a daily basis, some of them mentioned a change in motivation over time, "… through the last couple of years, it's then pretty well under control. So I haven't really gone out to look for information." Most commonly searched for topics were weight loss, recipes, explanations of symptoms, side effects of medications.

*Information Processing.* Information processing concerns the means by which participants make sense of collected information. Information sharing was found to be a way of processing information. Some participants shared information with healthcare providers to verify information. In one participant's words, "[you] find information through a variety of different sources…process it yourself, see what might work for you, and then take that to your physician or your nutritionist and…work together with them." Some participants reported gaining satisfaction from sharing information or strategies that they found effective with fellow diabetics and family members. Another aspect of information processing is resolving conflicts. Participants described different means of resolving conflicts, such as, (i) consulting with healthcare providers, (ii) comparing with existing verified knowledge, and (iii) by evaluating outcomes from following advice they were initially unsure of. With respect to validating information, one participant said they were able to determine the utility of a piece of information by looking at "...my results. Whether that be my readings, by my testing, by my scale, by how I'm feeling."

*Information Utilization.* This theme captures how information assisted participants with decision-making and affected their lifestyle. All participants reported that being diagnosed with T2DM and educating themselves about their disease significantly changed their lifestyle. Many participants reported positive outcomes as a result of better disease management, such as weight loss, increased activity level, feeling empowered and gaining confidence. As one participant said, "I feel very very good, very proud of myself. I know had I not been diagnosed I wouldn't have made those change[s]." Another aspect of information usage is personal autonomy. Some participants stated that they liked to be able to set their own course and make their own decisions with respect to information.

*Trust*. The theme of trust addressed participants' concerns with the validity of the information they found. This theme appeared in most of the interviews and emerged as a major concern. Some of the trusted sources that were mentioned by different participants are professional health care providers, verified commercial websites and educational/university resources. One participant said, "It has to come from a proper source. It has to come from the medical community or a medically backed community or…a university." Some participants mentioned trusting information that came from other diabetics from their social or online circles. Most of the participants mentioned distrusting online information that was not from sources they deemed as authentic. Some measures of authenticity of online resources that were mentioned are (i) rating and review of a website by other diabetics (ii) whether the website resources were approved by professional health care providers, and (iii) whether the information seemed to be neutral or sponsored by a group and therefore biased.

*Personalization.* Personalization refers to participants' need to receive information that is tailored to each individual. Five of the eight participants described how they felt the degree to which it is important to have personalized information changed over their disease trajectories. Many participants echoed ideas that "everybody [is] different." One participant said, "There can be a million type 2 [diabetics], but there'll be a million different ways for each individual to manage their type 2 [diabetes]." Another participant reported discovering that the instructions provided by physicians and dietitians were just "sort of guidelines" and that "each person has to tailor it to whatever their bodies can stand." Some participants also reported that personalization plays an important role in conflict resolution and determining whether or not they can trust an information source.

**Discussion**

**Summary of Key Findings.** This study aimed to better understand how patients with T2DM seek information, how they process information, and how they integrate information in their self-management routines. This process is represented in Figure 1. Patients may be motivated to search for information and in doing so, they may interact with healthcare or social networks. As they gather information they must then process that information, which includes resolving conflicts, evaluating trustworthiness of sources, and sharing information along the way. As they move from processing to utilizing information, patients personalize the information to fit their lives, incorporating it by making behavioral changes and using trial and error to determine whether or not that information is useful. If it is not, they may go through the cycle again to find new information that works for them. The information needs of patients may change over time, and they may reenter the cycle as needed. Although our interview guide was not focused on information utilization, this theme appeared multiple times in the participant narratives and warranted inclusion in our model.

**Comparison to Previous Research**. Past research indicated that active information seeking led to better adherence and health outcomes for T2DM patients,[5] and that barriers to self-education included lack of motivation and lack of personally relevant information.[16] These assertions were supported by our findings. Existing literature identified consulting with healthcare providers as an important component of processing.[9,11] Several of our participants agreed and added that interactions with peers and social networks were an important element of processing.

Some of our findings contradicted previous research. Past studies indicated that T2DM patients who use the internet would be younger and more newly diagnosed,[13] but our sample, comprised mostly of middle-aged adults who had had diabetes for several years, cited the internet as their most commonly used source. Some discussed being less active information seekers early in their diagnosis--it was only later that they gained the confidence and initiative to seek information on their own. One of the barriers to information seeking identified in the literature was a lack of clarity of content,[14] but our participants did not cite this as a problem. Instead, they identified assessing the trustworthiness of information sources and sorting through conflicting information from different sources as barriers to information processing. Finally, previous research indicated that lack of support from healthcare providers was a barrier to self education for T2DM patients,[16] however our findings indicated that lack of support from providers was actually a motivating factor that pushed people to look for alternate sources of information on their own. Many of the study participants reported the need to use personalized information for managing their T2DM. Although there are several existing works on personalizing medicine and/or treatment for T2DM patients using genetic or phenotypic interventions,[17] we did not find any overt references to existing literature that studied the utilization of personalized information in T2DM patients.

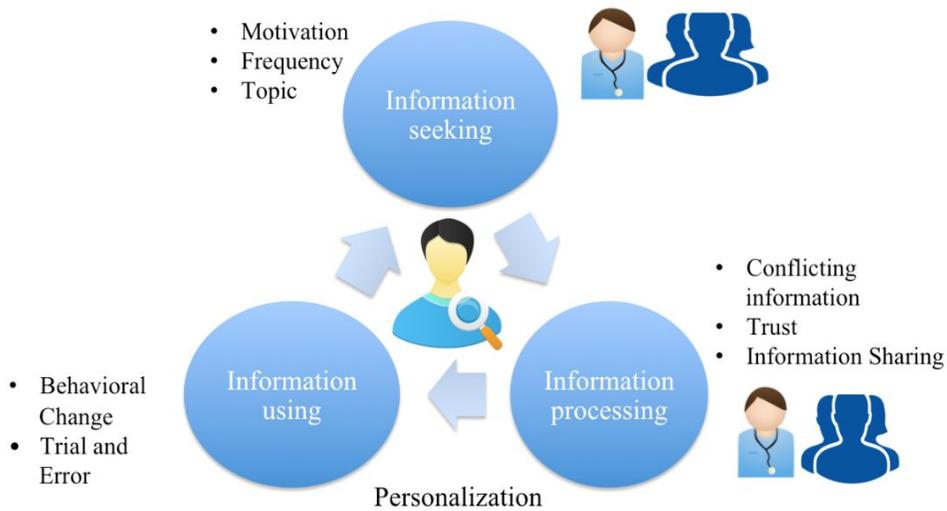

*Figure 1.* Summary of Key Findings

**Comparison to Existing Theories.** Although we used an inductive approach to design the study and analyze the results, we found that our data aligned with existing theories. The narratives of our participants echoed themes in the decisional balance model of behavior change.[18] This model captures the decisional balance of an individual's readiness to make a behavioral change. The pros of adopting healthy behavior are low in the early stages of change and increase as an individual progresses through different stages of change. In contrast, the cons of adopting healthy behavior are high in the early stages of change and decrease as an individual progresses through different stages. Some participants reported that they were initially overwhelmed by the work needed to manage their diabetes and found the guidelines they were given to be too generic--these were perceived cons of adopting healthy behaviors. However with time, they adjusted to the challenges and changes required to manage their condition and perceived less cons in healthy behaviors that they had adopted. For example, one participant reported feeling frustrated with the dietary restriction (restricting amount of carbohydrates for each meal) imposed on them after T2DM diagnosis. As they discovered that managing their diet was less difficult and restrictive than originally expected, the participant felt that managing T2DM became easier. They also reported experiencing unexpected benefits from managing their disease, such as weight loss, increased fitness, and higher self- esteem and efficacy.

Another theory that aligns with our results is the theory of planned behavior.[19] This theory proposes behavior intention as the best predictor of planned behavior and identifies three main factors that influence intention. These factors are attitude towards the behavior, subjective norms and perceived behavioral control. Several participants reported that the shared experiences and feedback from their peers influenced them to make behavioral changes (e.g., "if he/she can do it, so can I"). These social influences caused participants to view information coming from other diabetics as trustworthy and personalized, since the information was being derived from the lived experiences of someone they could relate to. Through the influence of their peer networks, subjective norms affected the behavior intention of our study participants. In addition, some participants described ways in which perceived control affected their behavioral intentions. For example, participants reported adhering to information or guidelines that are generated from sources that they found to be comprehensible, trustworthy, and accessible. One participant reported turning to her family whenever she has information needs regarding managing her T2DM, as they are doctors and she can access them frequently. Two participants described how switching from apathetic care providers to empathetic, concerned care providers motivated them to change their lifestyle for the better, which affected disease management. None of our participants specifically mentioned experiences that connect attitudes towards behavior with behavior intention. However, some participants mentioned how their attitude towards a behavioral change shifted over the disease trajectory according to the decisional balance model of behavior change as mentioned above. Although the theory of planned behavior considers interactions between the three factors of behavior intention (attitudes towards behavior, subjective norms, perceived behavior control), we didn't find specific evidence of such interactions in the narratives of our study participants.

**Implications.** The results of this study indicate a need for more targeted, individualized information and support for type 2 diabetics. Several participants identified a lack of personalized information as a key factor in the inability to adhere to T2DM management guidelines, which led them to experience increased glucose levels, difficulty managing A1C levels, frustration and anxiety. Participants mentioned following trial and error based approaches to tailoring information according to their needs and physiological conditions. We believe that personalizing information in the information-seeking phase may be more effective. Receiving personalized, trust-worthy information can lead to reduced burden in information processing and better information utilization as pointed out by multiple study participants. To address the issues identified by participants, we propose a web-based platform to assist T2DM patients in seeking personalized, trustworthy information, with the following characteristics:

- Physiological feedback, for example, how blood glucose levels, heart rate or other vitals change with a new activity is an important part of personalized information. This data could be collected with wearable sensors such as portable glucose monitors and smartwatches. These tools could be used to ensure that personalized information provided by the platform is yielding the desired results.
- To reduce the cognitive burden involved in sifting through online forums for personalized information, our web platform will include advice from other T2DM patients from online forums who have similar physiological and clinical characteristics to the user. The intuition behind this is, patients sharing similar physiological and clinical characteristics may benefit from the same set of intervention. This could be done by using advanced text mining and machine learning techniques.
- Solicit continuous user feedback to constantly improve the platform, e.g., the effectiveness of the intervention, decisional balance of user, ease of use, etc. The platform can also keep track of user feedback on the effectiveness of an information source, i.e., if intervention provided from an information source proves to be effective, that source will be prioritized in future.
- A number of participants mentioned being skeptical of information coming from major institutions because of concerns about advertising and money. In order to improve trust in the authenticity of sources for such users, the platform can provide information about sponsorship and origin of information and allow them to make their own decisions.
- Occasionally, the platform can incorporate feedback from healthcare providers to verify information and provide advice. This will contribute to increase users' trust on the platform. Also, getting information reviewed by their professional healthcare providers can lead the users to feel a sense of support from their providers, another unmet need of T2DM patients as pointed out by multiple study participants.

**Limitations.** A major limitation of this study is the sample size, which was small due to the study design and allotted timeframe. Moreover, our study participants were recruited online and may represent more active patients, as such our web platform may be most suited to such T2DM patients. This limits the generalizability of our findings and proposed intervention.

Another limitation is sample bias. Recruitment relied on a convenience-sampling model and it is possible that individuals who chose to respond to participation requests represent more online-based active patients than the general population. Individuals who have had particularly positive or negative experiences with information seeking might be more willing to discuss their experiences and may represent polarized ends of the spectrum. Similarly, patients who are satisfied with the information they passively receive from a healthcare provider might be less inclined to participate in this type of study. However, having an online-based sample makes this study more relevant to the promising research domain of digital health interventions. While it enabled us to understand the information seeking and processing behaviors of T2DM patients who are active online, the implications of the study can benefit the larger T2DM patient community as well. Because, establishing an accessible, personalized and trustworthy source of information is imperative to any T2DM patients irrespective of their reliance on online T2DM educational resources.

**Future Steps.** Some issues emerged that warrant further examination. Many participants mentioned dissatisfaction with formalized diabetic education programs. More research needs to be done on the effectiveness of these programs and potential ways to improve them to better meet the needs of newly diagnosed diabetics, as many of them are often directed towards these programs. Another recurring theme concerned the trustworthiness of web-based information sources. Many people rely on online resources for health information, so developing criteria to assess the authenticity of sources could be something patients would find useful.

**Conclusion**

This study aimed to better understand where patients with T2DM go for information, why they choose those sources, and how they integrate information in their self-management routines. Information seeking, processing, and utilization were characterized as cyclic behaviors, as patients are constantly locating, evaluating and utilizing information to manage their disease. Several key themes and areas for further research emerged from the data, such as, personalization of information, trustworthiness of information sources, and resolving conflicting information. Our research contributes towards finding interactions between the information seeking, processing and utilization behaviors of T2DM patients. This topic has not been studied extensively and our findings begin to fill in a gap in the existing literature.

**References**

1. More than 29 million Americans have diabetes; 1 in 4 doesn't know [Internet]. Centers for Disease Control and Prevention; 2014. Available from: http://www.cdc.gov/media/releases/2014/p0610-diabetes-report.html
2. Mokdad AH, Ford ES, Bowman BA, Dietz WH, Vinicor F, Bales VS, et al. Prevalence of Obesity, Diabetes, and Obesity-Related Health Risk Factors, 2001. JAMA. 2003 Jan;289(1):76–9.
3. Hilliard M, Sparling K, Hitchcock J, Oser T, Hood K. The Emerging Diabetes Online Community. Current Diabetes Reviews. 2015;11(4):261–72.
4. Chen G-D, Huang C-N, Yang Y-S, Lew-Ting C-Y. Patient perception of understanding health education and instructions has moderating effect on glycemic control. BMC Public Health. 2014 Apr;14(1).
5. Norris SL, Smith J, Schmid CH, Engelau MM, Lau J. Self-management education for adults with type 2 diabetes. Diabetes Care . 2002;25(7):1159–71.
6. Urbanski P, Wolf A, Herman WH. Cost-Effectiveness of Diabetes Education. Journal of the American Dietetic Association. 2008;108(4).
7. Jamal A, Khan SA, Alhumud A, Al-Duhyyim A, Alrashed M, Shabr FB, et al. Association of Online Health Information–Seeking Behavior and Self-Care Activities Among Type 2 Diabetic Patients in Saudi Arabia. Journal of Medical Internet Research. 2015Dec;17(8).
8. Kalantzi S, Kostagiolas P, Kechagias G, Niakas D, Makrilakis K. Information seeking behavior of patients with diabetes mellitus: a cross-sectional study in an outpatient clinic of a university-affiliated hospital in Athens, Greece. BMC Research Notes. 2015;8(1):48.
9. Longo DR, Schubert SL, LeMaster BA, Williams CD, Clore JN. Health information seeking, receipt, and use in diabetes self-management. Annals of Family Medicine. 2010;8(4):334–40.
10. Zare-Farashbandi F, Rahimi A, Lalazaryan A, Zadeh A. How health information is received by diabetic patients? Advanced Biomedical Research. 2015;4(126).
11. Low LL, Tong SF, Low WY. Social Influences of Help-Seeking Behaviour Among Patients With Type 2 Diabetes Mellitus in Malaysia. Asia-Pacific Journal of Public Health. 2015;28(1 Suppl).
12. Weymann N, Harter M, Dirmaier J. Quality of online information on type 2 diabetes: a cross-sectional study. Health Promotion International. 2014;30(4):821–31.
13. Lui C-W, Coll JR, Donald M, Dower J, Boyle FM. Health and social correlates of Internet use for diabetes information: findings from Australia's Living with Diabetes Study. Australian Journal of Primary Health. 2015;21(3):327.
14. Praveen PA. Trust and sources of health information –A study on diabetes mellitus in urban and rural areas of Thiruvananthapuram district,Kerala [dissertation]. [Thiruvananthapuram, Kerala]; 2010.
15. Valdez RS, Guterbock TM, Thompson MJ, Reilly JD, Menefee HK, Bennici MS, et al. Beyond Traditional

Advertisements: Leveraging Facebook's Social Structures for Research Recruitment. Journal of Medical Internet Research. 2014;16(10).
16. Crowe M, Whitehead L, Bugge C, Carlyle D, Carter J, Maskill V. Living with sub-optimal glycaemic control: the experiences of Type 2 diabetes diagnosis and education. Journal of Advanced Nursing. 2016
17. Raz I, Riddle MC, Rosenstock J, Buse JB, Inzucchi SE, Home PD, et al. Personalized Management of Hyperglycemia in Type 2 Diabetes: Reflections from a Diabetes Care Editors' Expert Forum. Diabetes Care. 2013;36(6):1779–88.
18. Prochaska JO, Velicier WF, Rossi JS, Goldstein MG, Al E. Stages of change and decisional balance for 12 problem behaviors. Health Psychology. 1994;13(1):39–46.
19. Ajzen I. The theory of planned behavior. Organizational Behavior and Human Decision Processes. 1991;50(2):179–211.